\documentclass[twocolumn,aps,prd,showpacs]{revtex4}
\usepackage{graphicx}
\usepackage{natbib}
\usepackage{amsbsy,amssymb,amsmath}
\usepackage[usenames]{color}
\usepackage{psfrag}
\usepackage[colorlinks,bookmarks]{hyperref}
\definecolor{linkblue}{rgb}{0,0,0.8}
\definecolor{linkgreen}{rgb}{0,0.5,0}
\hypersetup{pdfpagemode=UseNone, pdfstartview=FitH, linkcolor=linkblue, %
            citecolor=linkgreen, urlcolor=linkblue}

\newcommand{\beq}{\begin{equation}}
\newcommand{\eeq}{\end{equation}}

\def\vlg{v_{\rm LG }}
\def\vr{v_{\rm r}}
\def\zout{z_{\rm out}}
\def\zin{z_{\rm in }}

\def\ld{\left}
\def\rd{\right}
\def\fr{\frac}

\def\lcdm{$\Lambda$CDM}

\begin{document}

\title{Spatial and temporal tuning in void models for acceleration}

\author{Simon Foreman}
\email{sforeman@phas.ubc.ca}
\affiliation{Department of Physics and Astronomy, %
University of British Columbia, %
Vancouver, BC, V6T 1Z1  Canada}

\author{Adam Moss} 
\email{adammoss@phas.ubc.ca}
\affiliation{Department of Physics and Astronomy, %
University of British Columbia, %
Vancouver, BC, V6T 1Z1  Canada}

\author{James P. Zibin} 
\email{zibin@phas.ubc.ca}
\affiliation{Department of Physics and Astronomy, %
University of British Columbia, %
Vancouver, BC, V6T 1Z1  Canada}

\author{Douglas Scott} 
\email{dscott@phas.ubc.ca}
\affiliation{Department of Physics and Astronomy, %
University of British Columbia, %
Vancouver, BC, V6T 1Z1  Canada}

\date{\today}

\begin{abstract}
There has been considerable interest in recent years in cosmological models in which we inhabit a very large, underdense void as an alternative to dark energy. A longstanding objection to this proposal is that observations limit our position to be very close to the void centre. By selecting from a family of void profiles that fit supernova luminosity data, we carefully determine how far from the centre we could be. To do so, we use the observed dipole component of the cosmic microwave background, as well as an additional stochastic peculiar velocity arising from primordial perturbations. We find that we are constrained to live within 80 Mpc of the centre of a void---a somewhat weaker constraint than found in previous studies, but nevertheless a strong violation of the Copernican principle. By considering how such a Gpc-scale void would appear on the microwave sky, we also show that there can be a maximum of one of these voids within our Hubble radius. Hence, the constraint on our position corresponds to a fraction of the Hubble volume of order $10^{-8}$. Finally, we use the fact that void models only look temporarily similar to a cosmological-constant-dominated universe to argue that these models are not free of temporal fine-tuning.
\end{abstract}

\pacs{98.80.Es, 95.36.+x, 98.65.Dx}

\maketitle

\section{Introduction}

Modern observational data suggest that the standard model of cosmology (SMC; see, e.g., \cite{scott-smc}) now rests on firm foundations. Two of its main postulates are the Copernican principle and the existence of dark energy that closely resembles a cosmological constant. Nevertheless, inhomogeneous models that propose to do away with both of these ideas have been studied as alternatives to the SMC for over a decade (see \cite{mzs-precision,biswas-82pg,clarkson-rad} and references within for recent studies, or \cite{enqvist} for a more general review). These proposals aim to dispose of the perceived inelegance of dark energy by appealing to the fact that we are usually restricted to observing the Universe along our past light cone.  In particular, an increasing expansion rate {\it in time\/} [the standard explanation for the redshift-luminosity distance relation of recent surveys of Type Ia supernovae (SNe)] could also be interpreted as an expansion rate that increases {\it towards us spatially\/} if the assumption of large-scale homogeneity were to be abandoned. Indeed, supposing that we live in a large ($\sim$Gpc scale) underdense region that is isotropic about us, it is possible to obtain an exact match to the redshift-luminosity distance relation of the SMC (see, e.g., \cite{yoo-fit}).

While this is an encouraging sign, there are several obstacles to a wholesale adoption of such a ``void'' model as a solution to the mysteries surrounding dark energy. Voids of sufficient size occur with negligible probability in our current picture of structure formation \cite{strucfor}, and must be nearly spherical to accord with the observed isotropy in the cosmic microwave background (CMB). Further constraints abound, in the form of baryon acoustic oscillations \cite{gbh-bao}, local estimates of the Hubble rate \cite{feb-hubble}, the kinematic Sunyaev-Zeldovich effect \cite{gbh-ksz}, spectral distortions in the CMB \cite{cs-ydist}, and possible future measurements of cosmic shear \cite{zms-letter,gbh-shear}.  The information contained in the full CMB power spectrum has been found to place very strong constraints on void models \cite{zms-letter,mzs-precision}.

A different kind of constraint to consider is the maximum distance an observer could be from the centre of a void to match current observations. A rough guideline can be obtained by calculating the anisotropic redshift-luminosity distance relation that would result from an off-centre position \cite{bm-dipole,aa-sne}. However, it is well-known \cite{moffat95,Nakao:1995kw,tomita96,Humphreys:1996fd} that an off-centre observer would observe a prominent dipole component of the CMB, since photons traveling from the last scattering surface (LSS) would experience direction-dependent degrees of redshifting.

Previous studies \cite{aa-dipole,bm-dipole,biswas-82pg,ksi-dipole} have asked how far from the centre an observer could live before detecting a dipole exceeding the currently measured value. Our primary aim is to perform a thorough calculation of this distance, distinct in several ways from earlier estimates. First, we examine a broad subset of the space of possible void profiles, building upon the framework established in Refs.\ \cite{zms-letter,mzs-precision}. Second, we use the CMB dipole measured for the Local Group ($\vlg$) in obtaining our constraint, rather than, say, the dipole for the solar system, for reasons discussed below. Finally, we add an extra stochastic peculiar velocity component to the dipole induced by an off-centre position, arising from the perturbed velocity field created by density fluctuations at early times.

We begin in Sec.\ \ref{sec:form} by reviewing the formalism used in our investigation. In Sec.\ \ref{sec:multiplevoids}, we present computations demonstrating that no other large voids can be present within our Hubble volume. Thereafter, we describe the computational methods employed, and state our results for both radial (Sec.\ \ref{sec:compradcon}) and volume (Sec.\ \ref{sec:compvolcon}) constraints. We also find in Sec.\ \ref{sec:time} that the redshift-luminosity distance relation for supernovae imposes some fine-tuning of the {\it time\/} at which a void is observed. We conclude in Sec.\ \ref{sec:conc} with a summary of our work.

\section{Formalism}
\label{sec:form}

\subsection{Induced dipole}
\label{sec:ana}

Following standard practice, we consider the void to be described by a Lema\^{i}tre-Tolman-Bondi (LTB) spacetime \cite{ltb1,ltb2,ltb3}: an exact, spherically-symmetric dust solution to Einstein's equations.  Details of this solution and our numerical implementation of it are provided in Ref.~\cite{mzs-precision}. We first wish to calculate the CMB dipole induced by the void at an off-centre position, and then incorporate an additional stochastic peculiar velocity. The induced dipole is found using the method described in \cite{mzs-precision}. Briefly, the redshifts for two radial geodesics are considered, one incoming from the LSS (at $\zin$, set to 1000) to the observer, and the other outgoing through the void centre in the opposite direction ($\zout$).  The outgoing ray is constrained to originate at the same proper time (and hence the same matter density) as the incoming ray.

Setting $c=1$, the resulting dipole velocity $\vr$ can be found via
\begin{equation}
\label{v_r_eq} \vr = \frac{\zout-\zin}{2+\zout+\zin}.
\end{equation}
This determines the temperature anisotropy through \beq \fr{\Delta T(\theta)}{T} = \vr\cos\theta, \label{DTtheta} \eeq or, equivalently, the dipole amplitude via \beq |a_{10}| = \sqrt{\fr{4\pi}{3}}\vr. \eeq In general we expect higher multipoles to appear at higher order in $\vr$, but calculations involving the full non-radial geodesics confirm that the dipole contribution is dominant at the distances from the centre that we consider \cite{aa-dipole}. Numerically, our technique is found to agree with the analytical approach given in \cite{ksi-dipole} to better than one percent.

Note that our calculation of $\vr$ is equivalent to a calculation of the relative tilt between the comoving matter and radiation frames at the off-centre observation point.  Importantly, the LTB solution assumes a purely dust source, and hence cannot properly take into account the effect of radiation on the background evolution.  However, at late times the radiation density is much smaller than the matter density, and hence it is a good approximation to treat the radiation as a test field on the LTB spacetime, as we have done.  At early times (or sufficiently far outside the void), the void spacetime approaches homogeneous Friedmann-Lema\^itre-Robertson-Walker (FLRW), since we only consider growing mode void profiles.  Thus, again, in this regime we do not require the full relativistic solution including radiation and matter in spherical symmetry.

\subsection{Stochastic peculiar velocity}

Due to the complexity of describing perturbations on general LTB backgrounds, directly evolving the perturbed velocity to find the extra stochastic component of the dipole is not currently feasible. However, following the approach of \cite{mzs-precision}, we note that close to the void centre, shear is small enough that perturbations of matter density and expansion evolve essentially as in an open FLRW model. Thus, for the small scales making the dominant contribution to the dipole, we are able to apply the tools of Newtonian perturbation theory on FLRW backgrounds: for a single spatial component $v_i$ of the random peculiar velocity, the power spectrum reads \cite{llnew} 
\begin{equation}%
\mathcal{P}_{v_i}(k) = \frac{(aHf)^2}{3k^2} \mathcal{P}_\delta(k),
\end{equation}
where $f\simeq\Omega_{\text{m}}^{0.6}$ is a factor accounting for the suppression of growth of velocity fluctuations, and $\mathcal{P}_\delta(k)$ is the power spectrum for the matter fluctuations, $\delta \equiv \delta\rho/\rho$, at the centre of the void. For any observer, each $v_i$ will be normally distributed with mean zero and variance obtained by integration of the above spectrum:
\begin{align}
\nonumber \sigma^2 &= \int \mathcal{P}_{v_i}(k) \frac{dk}{k} \\%
&= \frac{8\pi G}{9} \rho_{\text{c}} \Omega_{\text{m}}^{1.2} \! \int_0^{k_{\text{LG}}} \mathcal{P}_\delta(k) \frac{dk}{k^3}, \label{sigmavPS}
\end{align}
where $\rho_{\text{c}}$ is the critical density, the FLRW scale factor $a$ is set to unity today, and the integral is cut off at the scale of the Local Group, $k\simeq 1.0$ $\text{Mpc}^{-1}$ (the integral is only weakly sensitive to the precise cutoff value).

We obtain $P_\delta(k)$ from the public Boltzmann code CAMB \cite{camb} by using parameters corresponding to an effective open FLRW model that gives the same physics at recombination and evolution along the central worldline as the void model (see \cite{mzs-precision} for details).  Importantly, in this approach we must assume that the primordial amplitude at the centre of symmetry is the same as that at the LSS, which is directly constrained by the CMB.  Although we do not have an explicit model for the formation of such a large void, we only consider growing mode void profiles in this work, so that at early times the void can be accurately described by linear theory on FLRW.  Thus, our assumption states that we can consider the standard inflationary primordial spectrum to be superposed on the early linear void.  Note also that Eq.~(\ref{sigmavPS}) agrees with the dominant part of a calculation of the dipole using linearized general relativity \cite{zs-gauge}.

The observed dipole will then be the sum of induced and stochastic peculiar velocities. We can treat this as a vector sum, letting one component of the stochastic velocity, say $v_1$, lie along the direction of $\vr$, with the other two components orthogonal.  Then, the observed dipole will have magnitude $v$ equal to the Euclidean norm of the vector ($\vr+v_1,v_2,v_3$). To derive the probability density function (pdf) for $v$, we recall that for a random variable $X$ with pdf $f_X(x)$ (the lower-case argument denoting a particular value of the corresponding upper-case random variable), the pdfs for $Y=X^2$ and $Z=\sqrt{X}$ are given respectively by \cite{grinstead-prob}
\begin{align}
f_Y(y) &= \frac{1}{2\sqrt{y}} \ld[f_X(\sqrt{y}) + f_X(-\sqrt{y})\rd], \label{distsq} \\
f_Z(z) &= 2z f_X(z^2). \label{distsqr}
\end{align}
With each $v_i$ normally distributed as above, applying (\ref{distsq}) to each component of ($\vr+v_1,v_2,v_3$) gives
\begin{align*}
f_{(\vr+v_1)^2}(x) &= \frac{1}{\sqrt{2\pi x}\sigma} e^{-(x+\vr^2)/(2\sigma^2)} {\rm cosh} \! \left( \frac{\vr \sqrt{x}}{\sigma^2} \right), \\
f_{v_2^2}(x) &= f_{v_3^2}(x) = \frac{1}{\sqrt{2\pi x}\sigma} e^{-x/(2\sigma^2)}.
\end{align*}
Adding the three squared components (which entails performing convolutions between their distributions) and applying (\ref{distsqr}) to the result yields the pdf for $v$, found to have the form of a ``skewed Maxwellian:''
\begin{equation}
\label{v_pdf_eq} f(v) = \sqrt{\frac{2}{\pi}} \frac{v}{\sigma \vr} e^{-(v^2+\vr^2)/(2\sigma^2)} \, \text{sinh} \! \left(\frac{\vr v}{\sigma^2}\right), \;\; v \geq 0.
\end{equation}
This can be integrated analytically to yield the cumulative density function for $v$ in terms of error functions, a lengthy expression that we omit here.

\subsection{Choice of measured dipole}

Before using a measured dipole to constrain our position, careful thought should be given to {\it which\/} measured dipole to use. References \cite{biswas-82pg,bm-dipole} use the dipole observed for the solar system's motion relative to the CMB sky (that is, the raw measured dipole after subtracting the motion of the detector relative to the Sun), $v \simeq 369$ km/s. Meanwhile, Ref.\ \cite{aa-dipole} uses $a_{10} = 10^{-3}$ ($v = 150$ km/s) and Ref.\ \cite{ksi-dipole} uses $a_{10} = 1.23 \times 10^{-3}$ ($v = 185$ km/s), possibly due to confusion about the relationship between $v$ and $a_{10}$. Since we employ a linear treatment of fluctuations, we should ignore nonlinear contributions to the dipole, which are important only on very small scales. Also, we must consider that the entire Local Group, which extends in space for roughly 1~Mpc, would be contained well within a void of the scale considered in this work. A second observer positioned in another galaxy within this group would therefore be expected to obtain the same limit as us as to how far from the void centre they could be. For these reasons, we choose the dipole measured for the Local Group itself, $\vlg \simeq 627$ km/s \cite{scottsmoot}, as the constraining value in our study.

\section{Computational methods and results}
\label{sec:comp}

\subsection{Multiple voids}
\label{sec:multiplevoids}

As an accessory to the constraint in Sec.~\ref{sec:compvolcon} on habitable volume within a void, we first consider whether there could be other large voids between us and the LSS. If so, they could be expected to leave prominent scars on the CMB sky. This has previously been examined for voids that do not attempt to mimic the SN evidence for acceleration (see, e.g., \cite{arnau-sky,bacc-sky,masina-sky,sakai-sky}), and also for so-called ``swiss cheese" models \cite{wessel-sky}, but our analysis will take place within a slightly different framework. A spherical void would produce an axially symmetric anisotropy pattern on the sky, with a profile determined by the void's radial profile, and with an angular diameter dependent on the void's size and distance from us. The anisotropy would be strongest in the centre of the pattern, corresponding to photons that travel from the LSS through the void centre. We calculated the temperature anisotropy for photons passing radially through a void in Sec.~\ref{sec:ana}, and we can reuse that technique here, but now placing the observer at distances from the void as great as the radius of last scattering, $r_{\text{LSS}}$, which is the maximum distance at which a void could be visible. In the parlance of Sec.~\ref{sec:ana}, the ``outgoing'' ray passes through the void, while we use the ``incoming'' ray to define the reference temperature $T$ as a matter of convenience.

\begin{figure}
\centering \mbox{\resizebox{0.45\textwidth}{!}{\includegraphics[angle=0,trim=20 20 10 10]{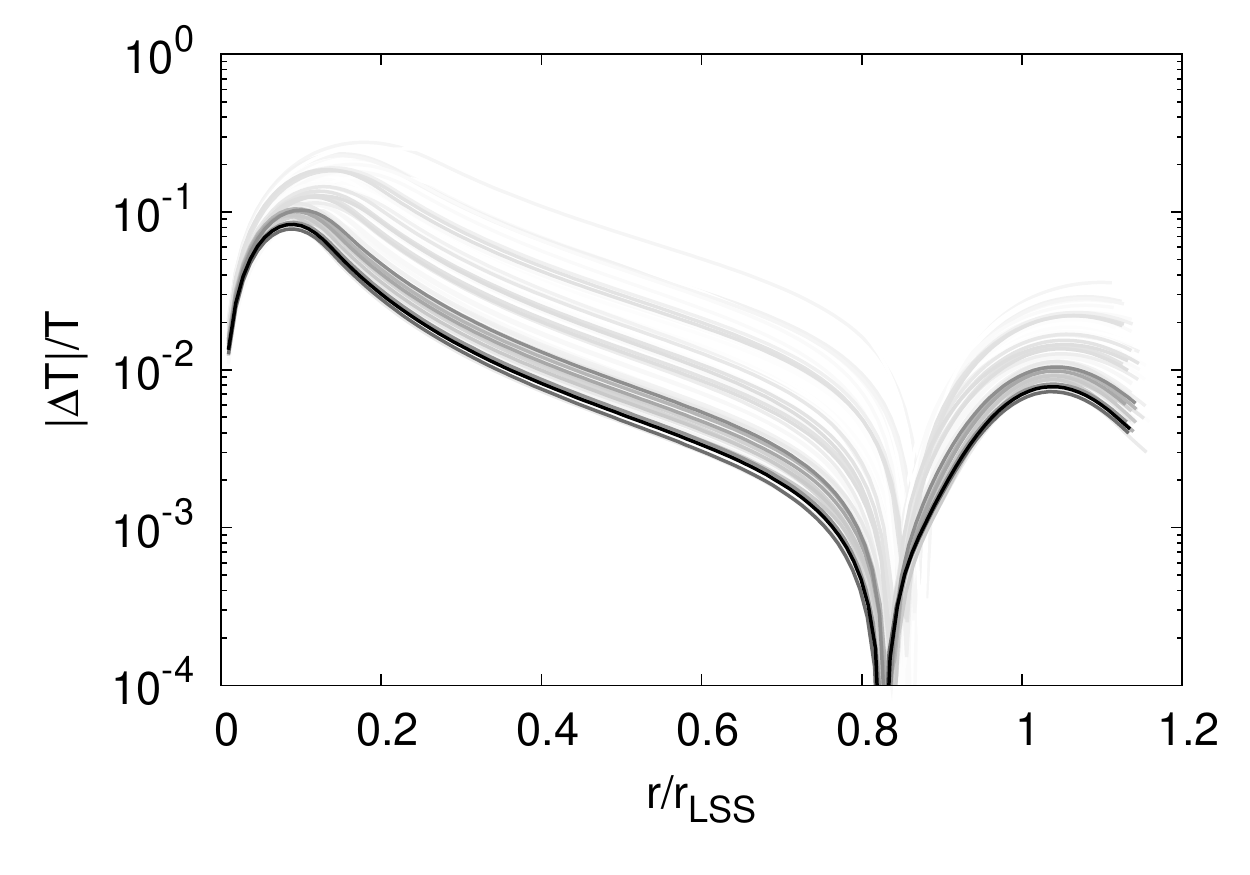}}} \caption{Maximum temperature anisotropy, $|\Delta T|/T$, vs.\ radial coordinate of the void centre for a selection of profiles from the chain (with no outer shells), with the grayscale level indicating the relative likelihood of the fit to CMB + SN data.  Apart from where $\Delta T$ changes sign at $r/r_{\text{LSS}} \simeq 0.8$, the magnitude of the effect always exceeds $10^{-3}$.} \label{deltt_fig}
\end{figure}

While the voids examined in \cite{mzs-precision} required an outer overdense shell to match CMB data, this would create a much larger effect than for just a central void.  Therefore, for a conservative estimate of the effect of other voids, we discard the shell for this calculation, leaving only the central underdensity in an Einstein-de Sitter (EdS) background. A plot of the resulting amplitude of the temperature anisotropy, $|\Delta T|/T$ [calculated using Eqs.~(\ref{v_r_eq}) and (\ref{DTtheta})] for a selection of void profiles from our Markov chain (described in Sec.~\ref{sec:compradcon}) is shown in Fig.~\ref{deltt_fig}, with the grayscale level indicating the likelihood of the fit to CMB + SN data. Up until $r/r_{\text{LSS}} \simeq 0.2$, the local dipole effect described in Sec.~\ref{sec:ana} dominates, while past $r/r_{\text{LSS}} \simeq 0.8$, the void intersects the observer's LSS, resulting in the regular Sachs-Wolfe effect. Between these regions, a non-linear integrated Sachs-Wolfe (ISW), or Rees-Sciama, effect contributes a $|\Delta T|/T$ with magnitude greater than $10^{-3}$. In comparison, the WMAP Cold Spot has roughly $|\Delta T|/T \simeq 4 \times 10^{-5}$ (see, e.g., \cite{coldspot}), so the presence of other large voids (i.e., of the kind required locally to fit the SN data) within our observable volume is ruled out by observations. In a recent study, no evidence was found for a void of radius $\sim200 h^{-1}$ Mpc as a possible source for the Cold Spot itself \cite{bremer-coldspot}.

\subsection{Radial constraint}
\label{sec:compradcon}

We use COSMOMC \cite{cosmomc} to generate Markov-Chain Monte-Carlo chains of void models as described in \cite{mzs-precision}, distributed according to fits to the full CMB temperature power spectrum (using data from WMAP7 \cite{wmap1,wmap2}, ACBAR \cite{acbar}, BOOMERANG \cite{boomerang}, CBI \cite{cbi}, and QUAD \cite{quad}) and SNe Ia (adopting the Union2 compilation \cite{union2}, which covers the range $z=$ 0.015--1.4). Each model is parametrized by its radial density profile $\delta(t,r)$ using a sum of polynomials representing a central underdensity and an overdense shell located at $z>1.5$ (see \cite{mzs-precision} for explicit expressions), and approaches an EdS background at large distances.  These models consist of pure growing mode profiles, and hence are specified by a single radial profile.

At a given proper distance from the centre (measured along a comoving slice), we evaluate the local dipole induced by each void model using Eq.\ (\ref{v_r_eq}), and then, using Eq.\ (\ref{v_pdf_eq}), calculate the probability that the observed dipole $v$ (which includes the stochastic peculiar velocity) will be less than $\vlg$. The resulting value is checked by building a normalized histogram from samples of the distribution (\ref{v_pdf_eq}) and integrating up to $\vlg$. Averaging the probabilities over the entire chain of models, and repeating this process for different physical distances, allows us to obtain confidence limits on how close to the centre of the void an observer must be in order to see a dipole less than $\vlg$.

\begin{figure}[t]
\centering \mbox{\resizebox{0.45\textwidth}{!}{\includegraphics[angle=0,trim=20 5 10 10]{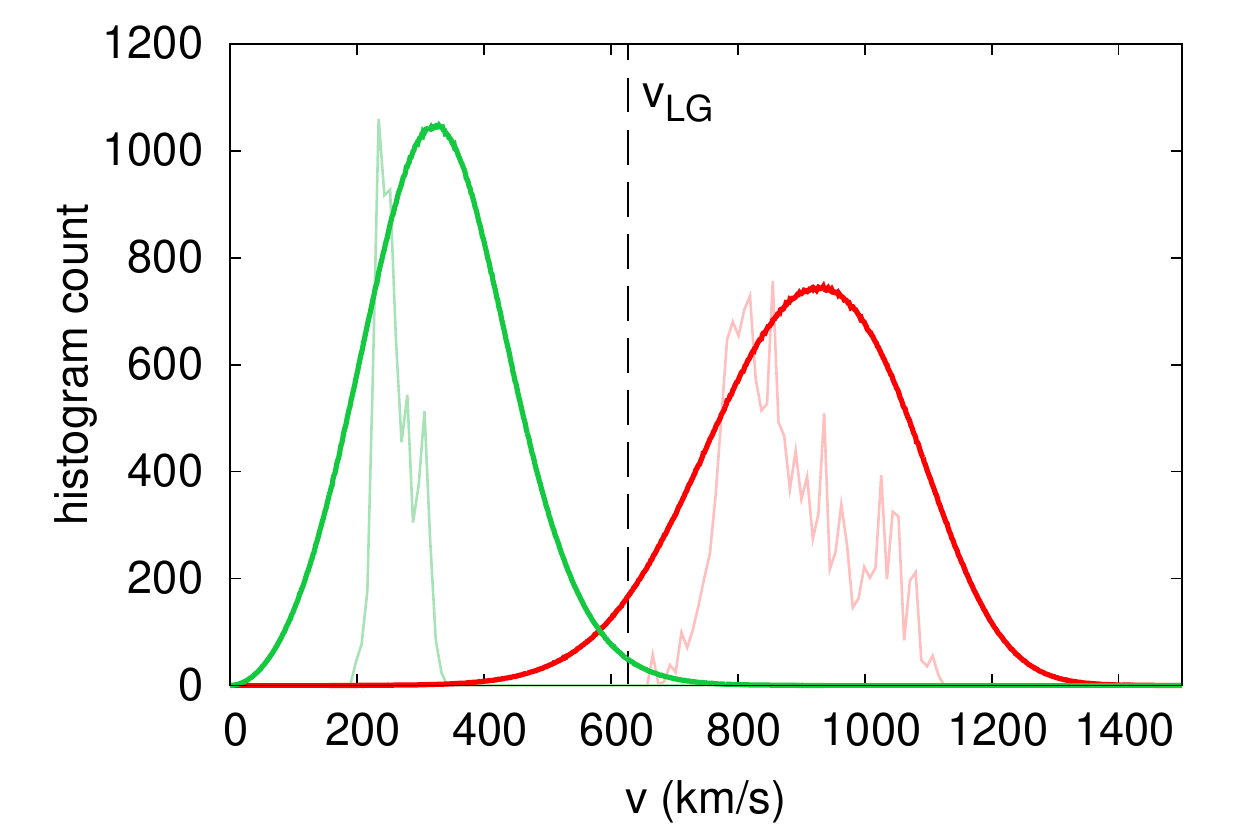}}} \caption{Distribution of dipole magnitude $v$ at radii 24 Mpc (green curves) and 80 Mpc (red curves) over the chain of void models, with random peculiar velocity (solid colours) and without (faint colours, apparently noisy due to the finite Markov chain sampling). At each distance, curves are scaled to the same height. The addition of peculiar velocity is seen to significantly widen both distributions.} \label{hist_comp_fig}
\end{figure}

We find that an observer must be within 80 Mpc of the centre of a void in order to observe a dipole less than $\vlg$ with greater than 95\% confidence. The loosening of the constraint on radial distance in comparison with past studies is caused by the three additional considerations described above: our broad exploration of the space of profiles; our choice of the Local Group dipole instead of a value based on smaller-scale motion to use as the empirical value; and the inclusion of the extra stochastic peculiar velocity component. If, for example, we use the dipole from \cite{aa-dipole}, namely 150 km/s, then the third consideration alone results in a radial constraint of 24 Mpc, pictured in green in Fig.~\ref{hist_comp_fig}, while the constraint using $\vlg$ is seen in red. Despite being significantly looser than previous estimates, this new limit on radial distance using the CMB dipole anisotropy is still much tighter than that imposed by modern supernovae surveys \cite{bm-dipole,aa-sne}.

\subsection{Volume constraint}
\label{sec:compvolcon}

In addition to endeavouring to limit an observer's distance from the centre, it is also worthwhile to ask what fraction of the total volume of a void one could fall within and still see a dipole less than $\vlg$, since this fraction takes into account the physical extent of the void itself. However, we must also consider the prior probability of finding ourselves within a void at all, and this depends on whether other voids of a suitable size can exist between our current position and the LSS.

As detailed in Sec.~\ref{sec:multiplevoids}, there can be a maximum of one void within our Hubble radius (namely, the central void which can fit the SN data).  Therefore, we should actually seek the fraction of our {\it entire Hubble volume} we could reside in to give an appropriate dipole. If we imagine a process of placing an observer at a uniformly random position, this fraction can be interpreted as the probability that the observer would end up seeing a dipole that agrees with $\vlg$. It is more sensible to picture this process occurring at an early time, so we evaluate the corresponding volumes on a spacelike slice corresponding to the time of last scattering. The process for obtaining confidence limits on the volume proceeds similarly to the radial calculation, described in Sec.~\ref{sec:compradcon}.

The result is that an observer must inhabit a fraction of the Hubble volume of order $10^{-8}$, again in order to have $v < \vlg$ at 95\% confidence. This is a clear violation of the Copernican principle, since observations force our position within the observable universe to have been fine-tuned to about one part in 100 million at early times. If we live in a void, our position becomes very special indeed.

\section{Temporal position}
\label{sec:time}

In light of the constraint on {\it where} we must live within a void, it is natural to ask if there might also be some restrictions on {\it when} we live. It has often been pointed out that the SMC suffers from such temporal tuning, in that the densities of matter and dark energy are comparable today but begin to differ by many orders of magnitude in the relatively near past or future (see, e.g., \cite{carroll-coincidence}). Any void model has an inherent time dependence, since the radial density profile ``flattens out'' into the past while the central underdensity deepens and the outer shell becomes even denser into the future. Thus, it can be expected that observational relations should be drastically altered at different times.

\begin{figure}
\centering \mbox{\resizebox{0.45\textwidth}{!}{\includegraphics[angle=0,trim=20 5 10 10]{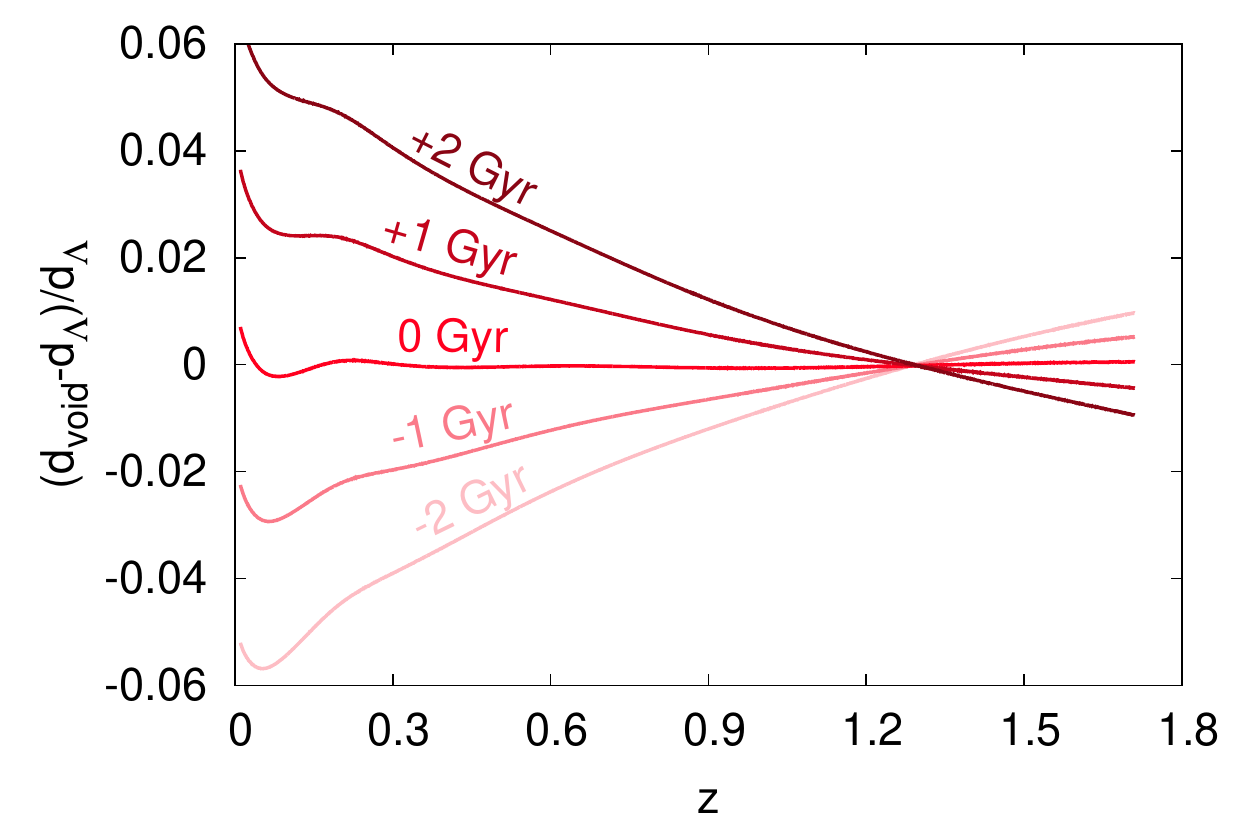}}} \caption{Fractional difference between luminosity distance in \lcdm\ and in a void whose redshift-luminosity distance relation matches well to \lcdm\ at the current epoch, plotted at different observation times. Both relations are recalculated at each observation time. The discrepancy exceeds 2\% at $\pm$1 Gyr, and increases at later or earlier times. The worse fit at low redshift stems from our constraint that the void profile be smooth at the centre.} \label{dlcomp_fig}
\end{figure}

It is possible to investigate this by calculating the redshift-luminosity distance relation for a void as seen by an observer at different times, and comparing it to that of the SMC (otherwise known as \lcdm) at those times. (We emphasize that relations in both void and \lcdm\ models are time-dependent, so both must be recalculated for each observation time.) In Fig.~\ref{dlcomp_fig}, we examine a void profile that matches \lcdm\ well at the current epoch, and find that the fractional difference between luminosity distances of the void and \lcdm\ increases for times other than the present, as expected.

\begin{figure}
\centering \mbox{\resizebox{0.45\textwidth}{!}{\includegraphics[angle=0,trim=20 5 10 10]{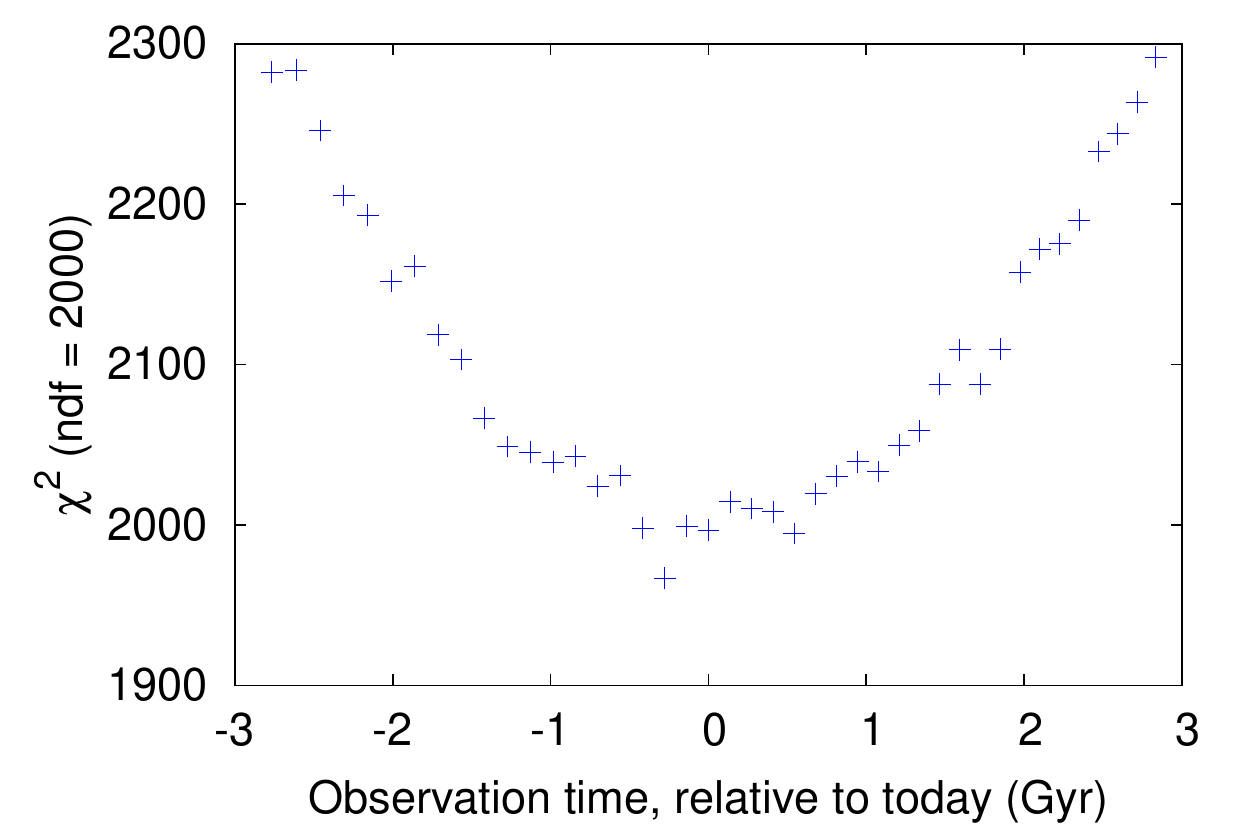}}} \caption{We simulate the $\mu$--$z$ relationship for \lcdm\ at various times given by 2000 SNe uniformly distributed from $z=$ 0.01--1.7, with a Gaussian dispersion characterized by $\sigma_{\mu}=0.15$. We then calculate $\chi^2$ values at those times for an effective ``fit'' of this relationship to that of a void profile, averaged over 30 fiducial data sets. There is a clear preference for the current epoch, with times more than $\sim$1.5 Gyr into the past or future exhibiting severe deviations from \lcdm .} \label{cs_mu_fig}
\end{figure}

Attempting to generate fiducial data can also lend insight into how observations in a void would change with time. To this end, we simulate 2000 SNe, uniformly distributed in redshift from $z=$ 0.01--1.7, by calculating the distance modulus $\mu$ for each one according to \lcdm\ at a certain time and adding on a realistic Gaussian dispersion with $\sigma_{\mu}=0.15$. We then find the $\chi^2$ of an effective ``fit," treating the $\mu$--$z$ curve at that time from the void profile used in Fig.~\ref{dlcomp_fig} as the null hypothesis to be compared with the simulated \lcdm\ data. This process is repeated at different observation times, and the results are shown in Fig.~\ref{cs_mu_fig}. In addition to the best fit occurring at the current time, we see that observing the void more than $\sim$1.5 Gyr into the past or future results in $\Delta \chi^2 \gg \sqrt{2000} \simeq 45$, giving a redshift-luminosity distance relation that noticeably deviates from the behaviour of \lcdm\ at that time.

Thus, a void that fits the relation of standard \lcdm\ does so only temporarily. It would therefore be rather coincidental to find ourselves living at a time when a void matches \lcdm\ so closely. Of course, better or worse simulated data would yield a more or less obvious temporal fine-tuning.  The important point, however, is that void models are not free of the temporal tuning that appears to plague \lcdm.

\section{Conclusions}
\label{sec:conc}

Our primary goal in this paper has been to obtain rigorous limits on the region an observer could inhabit within a cosmic void while detecting a CMB dipole less than the measured value. We have examined a wide range of void profiles for this purpose and clarified the best choice for the measured dipole. It is conceivable that a peculiar velocity in the proper direction could cancel a large dipole induced by a void, and we have accounted for that possibility in a probabilistic fashion by including an extra velocity distribution for the observer, derived from the expected primordial perturbation spectrum.

We found that an observer must be located within 80~Mpc of the void centre, or within a fraction of the Hubble volume of order $10^{-8}$, to see a dipole less than $\vlg$ at 95\% confidence. This constraint is weaker than previous estimates, both as a radius and as a fraction of the void radius (voids in our chain have average radii of roughly 3~Gpc, giving an average constraint of approximately $2.5$\% of the void radius, as opposed to the 1\% found in \cite{bm-dipole}). We also showed that there can be a maximum of one void large and deep enough to match the SN data within our Hubble volume by examining the anisotropy any other voids would create in the CMB. We additionally highlighted that the redshift-luminosity distance relation of a void can resemble that of \lcdm\ for only a brief epoch, and thus the time at which we observe a void is subject to some degree of fine-tuning.

These results indicate tension with the Copernican principle, severe in space but also present in time, as an unavoidable consequence of void models to explain cosmic acceleration. When coupled with the low probability of such a void forming by chance through the process that created the other matter perturbations \cite{strucfor}, it appears exceedingly unlikely that the Universe would arrange itself in such a configuration.

A natural question is whether departures from spherical symmetry, i.e.\ moving beyond the LTB model, could alleviate the spatial tuning we have examined here.  (This important question was addressed very recently in \cite{bs10}.)  Because a void required to fit the SN data must be Gpc-scale, void models with departures from a spherical profile on the order of Gpc might substantially reduce the spatial tuning problem, simply because the central position would not be well defined in such models.  However, nonlinear structures on Gpc scales (i.e.\ the features producing the departures from spherical symmetry) should produce an obvious nonlinear ISW imprint on the CMB, as we highlighted in Sec.~\ref{sec:multiplevoids}.  Therefore, while such models which depart significantly from spherical might lessen the tuning problem from the dipole, they would be expected to generate considerable power in higher multipoles of the CMB.  It appears inevitable that any departures from spherical must be on scales much smaller than Gpc, in which case the spatial tuning problem persists.

While these considerations may be seen as more philosophical than concrete, they complement recent thorough comparisons of void models with a wide variety of observational data that find strong discrepancies between the two \cite{mzs-precision,biswas-82pg} (however, see also \cite{clarkson-rad}). As an example, in \cite{mzs-precision} alone it is found that while voids can be made to fit SN and CMB data, they possess Hubble rates, ages, radial BAO scales, Compton $y$-distortions, and local matter fluctuation amplitudes that all come into conflict with observed values. There are many prospects as well for future constraints, including improved BAO measurements from surveys such as PAU \cite{pau} and BOSS \cite{boss}, more precise CMB spectra from Planck \cite{planck}, and perhaps even other limits on spatial position from cosmic parallax and redshift drift \cite{cosmicparallax,cosmicparallax2}. It is clear that void models continue to face serious obstacles to becoming successful alternatives to the SMC.


\section*{Acknowledgments}

This research was supported by the Natural Sciences and Engineering Research Council of Canada and the Canadian Space Agency.

\bibliography{void_dipole_v2}

\end{document}